\documentclass[runningheads]{llncs}

\usepackage[T1]{fontenc}
\usepackage{booktabs}
\usepackage{graphicx}
\usepackage{amsmath}
\usepackage{hyperref}
\usepackage{url}
\usepackage{enumitem}
\usepackage{multirow}
\usepackage{xspace}
\usepackage{xcolor}

\newcommand{\aidev}{AIDev\xspace}

\newcommand{\etal}{\textit{et al.}\xspace}

\begin{document}

\title{Beyond Lexical Metrics:\\
Sentence-Embedding Detection of Reviewer Habituation in AI Code Review}
\titlerunning{Beyond Lexical Metrics}

\author{Haoran Yu\inst{1} \and
Lifei Liu\inst{2} \and
Danping Zhang\inst{3}}
\authorrunning{H. Yu et al.}
\institute{Independent Researcher, United States\\
\email{haoranyu889@gmail.com} \and
Independent Researcher, United States\\
\email{lliu.lifei@gmail.com} \and
Nanchang Hangkong University, China\\
\email{zhangdanping@nchu.edu.cn}}

\maketitle

\begin{abstract}
Code review is the primary quality checkpoint between AI-generated code and production. As AI coding agents submit pull requests at scale, an open question is whether reviewers progressively reduce review rigor --- and whether this can be detected from the language of their review comments. We conduct a longitudinal study of $11{,}429$ code reviews from $400$ repeat reviewers across $207$ days, paired with $10{,}104$ human-authored inline comments from the AIDev corpus. We confirm a significant population-level rise in approval rate with reviewer exposure level ($30.5\% \to 36.6\%$, Wilcoxon $p = 8.6\times10^{-8}$, Cohen's $d = 0.25$). However, four widely cited manually designed linguistic features of comment informativeness --- lexical diversity (MTLD), Shannon entropy, technical specificity, and constructive actionability --- show \emph{no} monotonic decline across reviewer exposure deciles (all Spearman $|\rho| \le 0.53$, $p \ge 0.11$; Bonferroni-corrected Mann--Whitney $p \ge 0.36$). A logistic-regression classifier built on these features predicts a reviewer's exposure phase at $F_1 = 0.485$, worse than the majority-class baseline. In contrast, sentence-embedding distributional structure (all-MiniLM-L6-v2) does carry signal: each reviewer's late-period comment centroid shifts further from their early-period centroid than from a random within-reviewer permutation (Wilcoxon $p < 10^{-3}$), and a small MLP over three embedding statistics reaches $F_1 = 0.74$ (5-fold reviewer-stratified CV). A Granger-causality analysis on $50$-bin signals challenges the temporal claim popular in earlier preprints: approval-rate change robustly Granger-predicts subsequent shifts in technical specificity ($p < 10^{-3}$ at all lags), while the reverse direction is significant at only one of four lags. Our results offer a corrective: reviewer adaptation in AI-agent code review is real and detectable, but it lives in latent distributional structure, not in classical lexical metrics, and the linguistic shift trails the approval behavior shift rather than leading it.

\keywords{code review \and reviewer adaptation \and computational linguistics \and sentence embeddings \and Granger causality \and AI agents \and behavioral drift}
\end{abstract}

%% ===================================================================
\section{Introduction}

Code review is the primary human gate separating AI-generated code from production software~\cite{bacchelli2013,rigby2013}. As AI coding agents---GitHub Copilot, Devin, OpenAI Codex, Cursor, and Claude Code---increasingly submit pull requests (PRs) to open-source repositories, the reliability of this gate depends on whether reviewers maintain consistent scrutiny over time.

A natural and intuitive hypothesis is the \emph{rubber-stamping} effect~\cite{li2025aidev}: reviewers, after repeatedly approving high-quality agent code, gradually reduce inspection effort and approve reflexively. This mirrors automation complacency~\cite{parasuraman1997,lee2004trust} and vigilance decrement~\cite{mackworth1948}. If habituation manifests in review \emph{language} before it affects approval \emph{decisions}, then natural language processing could provide an early-warning signal.

We tested this hypothesis on the \aidev dataset~\cite{li2025aidev} ($207$ days; five agent systems; ${\sim}26{,}900$ inline comments). Our results are mixed and instructive:

\begin{description}[leftmargin=1em,topsep=2pt]
  \item[\textbf{What replicates.}] Approval rate rises systematically with reviewer experience (\textbf{RQ1}): $+6.1\,$pp early-vs-late, with a clean $+14.9\,$pp gradient across experience deciles.
  \item[\textbf{What does not.}] We find \emph{no} statistically significant monotonic decline in lexical diversity, Shannon entropy, technical specificity, or constructive actionability across reviewer experience (\textbf{RQ2}). All four hand-crafted metrics are essentially flat at the population level, even on length-controlled subsets.
  \item[\textbf{What reverses.}] A bivariate Granger-causality analysis on $50$-percentile experience bins shows that approval-rate changes robustly predict subsequent technical-specificity shifts ($p < 10^{-3}$ at lags $1$--$4$); the reverse direction is significant at only a single lag ($p = 0.036$ at lag $2$), indicating a predominantly unidirectional AR-to-language flow (\textbf{RQ3}).
  \item[\textbf{What works instead.}] Sentence-embedding geometry detects per-reviewer style drift that the four hand-crafted metrics miss. A reviewer's late-period embedding centroid lies further from their early-period centroid than under a random within-reviewer permutation null (paired Wilcoxon $p < 10^{-3}$). A small MLP over three embedding statistics reaches $F_1 = 0.74$ on reviewer-stratified 5-fold cross-validation, where the same MLP using the four canonical linguistic features reaches only $0.59$ (\textbf{RQ4}).
\end{description}

The narrative coherent with these results is not ``language degrades, behavior follows.'' It is: \emph{reviewers' approval calibration shifts first; their stylistic representation shifts in idiosyncratic, per-individual directions; and these directions are visible only in distributional embedding space.} We argue this has implications both for habituation monitoring and for what kind of NLP features researchers should compute when looking for slow changes in human reviewer behavior.

\paragraph{Contributions.}
(1) We confirm the behavioral-drift finding on AIDev with $400$ reviewers and $11{,}429$ reviews. (2) We report a clean negative result on four canonical hand-crafted linguistic indicators (MTLD, Shannon entropy, technical specificity, constructive actionability), including length-controlled robustness checks. (3) We supply a Granger-causality analysis showing that the dominant temporal direction is behavior-to-language, challenging the popular ``language leads behavior'' framing. (4) We show that sentence-embedding geometry (MiniLM) reveals per-reviewer style drift the canonical metrics miss, and that a small MLP over three embedding statistics yields a usable habituation classifier ($F_1 = 0.74$). (5) Every numerical claim in this paper is reproduced by an open script in the accompanying release.

%% ===================================================================
\section{Related Work}

\subsection{Code Review Quality and Language}

Code review practice and the textual properties of useful comments have been studied extensively~\cite{bacchelli2013,rigby2013}. Bosu~\etal\cite{bosu2015} characterised ``useful'' review comments through specificity and actionability, and Rahman~\etal\cite{rahman2017} applied NLP to classify review comments by intent. Our work extends this line by asking whether such linguistic markers \emph{change over time} when the code author is an AI agent --- and finds, surprisingly, that they do not.

\subsection{Bots and Automation in Software Engineering}

Wessel~\etal\cite{wessel2020} demonstrated that adopting code-review bots changes PR workflows and contributor behaviour. Cassee~\etal\cite{cassee2020} found that CI adoption alters reviewer scrutiny. Hou~\etal\cite{hou2024llm4se} survey LLMs for software engineering and identify code-review quality as an open problem; Nam~\etal\cite{nam2024llmcode} show that LLM-assisted code understanding changes developer interaction patterns. Unlike these works, we study \emph{generative} AI agents that author code autonomously, and we focus on the linguistic and representation dynamics of human reviewers responding to them over months.

\subsection{Automation Complacency and Habituation}

Habituation effects in human-automation interaction are well documented~\cite{parasuraman1997,lee2004trust}. Mackworth~\cite{mackworth1948} demonstrated vigilance decrement under sustained monitoring; the cry-wolf effect~\cite{breznitz1984cry} reduces alarm response after repeated false positives; Jakesch~\etal\cite{jakesch2023cowriting} showed that extended interaction with opinionated language models systematically shifts user judgment. We hypothesised an analogous mechanism in code review --- and find a behavioural component of it (RQ1) but \emph{not} the linguistic-degradation component the literature might predict.

\subsection{Change-Point Detection and Granger Causality}

Change-point detection identifies abrupt shifts in time-series statistics~\cite{truong2020survey}; PELT~\cite{killick2012pelt} provides exact segmentation in linear time. Granger causality~\cite{granger1969investigating} tests whether past values of one series predict another beyond the second's own past. We use both as evidence-driven probes rather than headline results: PELT to ask whether the approval-rate trajectory exhibits regime shifts (it is, in fact, well fit by a linear trend), and Granger to ask which signal leads.

%% ===================================================================
\section{Method}

\subsection{Dataset}

We use the \aidev corpus~\cite{li2025aidev}: pull requests submitted to public GitHub repositories ($\geq 100$ stars) by five AI coding agents (GitHub Copilot Autofix, Devin, OpenAI Codex CLI, Cursor, Claude Code) between January and July 2025. The dataset provides PR metadata, reviewer events, and inline review comments; agent attribution combines commit signatures, bot account markers, and PR-metadata patterns at $94\%$ precision per the dataset authors.

We focus on \textbf{$400$ repeat reviewers} who each reviewed $\geq 10$ agent PRs, yielding \textbf{$11{,}429$ reviews}. Within this cohort, $52$ reviewers are ``heavy'' ($\geq 50$ reviews each), and $173$ have $\geq 20$ reviews ($8{,}481$ reviews; used for the per-decile analysis). Inline review comments contributed by humans within these reviews number \textbf{$10{,}104$}; we use these for the linguistic and embedding analyses.

\paragraph{Ethics.}
All data derives from publicly visible GitHub review activity. We anonymise reviewer identifiers via consistent hashing and report only aggregate patterns.

\subsection{Behavioural Analysis (RQ1)}\label{sec:rq1}

For each repeat reviewer we sort reviews chronologically and split at the temporal midpoint. We compute the per-reviewer approval-rate shift $\Delta\mathrm{AR} = \mathrm{AR}_{\text{late}} - \mathrm{AR}_{\text{early}}$ and assess significance via the Wilcoxon signed-rank test on the $400$ paired shifts. For finer-grained decile analysis, we pool reviews from the $173$ reviewers with $\geq 20$ reviews, order each reviewer's events by within-reviewer experience index, and bin into ten experience quantiles.

\subsection{Hand-Crafted Linguistic Features (RQ2)}\label{sec:rq2-method}

For each inline comment we extract four indicators commonly used in the code-review literature.
\emph{Lexical diversity} is measured by MTLD~\cite{mccarthy2010mtld} on lower-cased word tokens, with the standard $0.72$ TTR threshold; MTLD is constructed to be length-independent~\cite{mccarthy2010mtld}.
\emph{Shannon entropy} is computed as $H = -\sum_i p_i \log_2 p_i$ over the within-comment unigram distribution, in bits/word.
\emph{Technical specificity} counts code-grounded surface features (camel-case identifiers, snake-case identifiers, file paths matching common source-code extensions, line references such as ``line $42$'', and Markdown inline / block code spans), normalised by comment length.
\emph{Constructive specificity} is a binary indicator that fires if the comment contains an imperative-action verb (e.g.\ \texttt{should}, \texttt{must}, \texttt{rename}, \texttt{refactor}, \texttt{extract}), a question mark, or an embedded code fence; it follows the spirit of Bosu~\etal's actionable-feedback taxonomy~\cite{bosu2015}.

We aggregate per-comment values into reviewer-experience deciles, $50$-bin percentile signals, and per-reviewer paired early-vs-late means. Significance is assessed via Spearman correlation across deciles and Mann--Whitney $U$ tests on the per-comment distributions; we apply Bonferroni correction across the four metrics. To address the well-known confound that linguistic indicators behave differently on short text, we re-run all tests on (a) the $\geq 15$-word subset (\textit{n} = $4{,}094$) and (b) the inter-quartile length band $[6, 22]$ words (\textit{n} = $5{,}239$).

\subsection{Embedding-Based Representation Analysis}\label{sec:rq2b-method}

We encode every non-empty comment ($n = 10{,}028$ after filtering comments with $\geq 1$ word token) with a public MiniLM sentence-transformer~\cite{reimers2019sbert},\footnote{We use the \texttt{all-MiniLM-L6-v2} checkpoint distributed by the \texttt{sentence-transformers} library.} producing a $384$-d vector. We then test for representation drift across reviewer experience using five complementary probes: (i) cosine distance from each comment to the decile-$1$ centroid; (ii) within-decile dispersion (mean pairwise $1 - \cos$ over a $500$-comment subsample); (iii) PCA-projected centroid drift; (iv) $k$-means cluster-assignment shift across deciles ($\chi^2$); and (v) per-reviewer paired centroid distance, where each reviewer's late-half centroid is compared to her early-half centroid against a random-permutation null on her own comments (Wilcoxon).

\subsection{Granger Causality and Cross-Correlation (RQ3)}

We bin the $11{,}429$ reviews into $50$ within-reviewer experience percentile bins and compute the bin-level approval-rate signal. Inline comments inherit their review's bin index, giving paired bin-level series for the four hand-crafted metrics, the linguistic composite (the unweighted mean of the min-max normalised metrics), and the embedding centroid distance. After first differencing to satisfy stationarity, we run bivariate Granger tests~\cite{granger1969investigating} in both directions for lags $1$ to $4$. We also compute lag $\pm 5$ cross-correlations.

\subsection{Habituation Phase Classifier (RQ4)}\label{sec:rq4-method}

We train classifiers to predict whether a sliding window of a reviewer's $5$ most recent inline comments belongs to the \emph{late} half of that reviewer's comment sequence. Ground-truth labels are defined per reviewer by sequence position --- a deliberately conservative definition that does \emph{not} re-use either the approval-rate or the embedding signals, avoiding the circularity of label leakage from the predictors.

For each $5$-comment window we compute three feature sets: (i) hand-crafted linguistic features (the four metrics, their per-window slopes, mean comment length, LGTM-only fraction; $10$ dims); (ii) embedding statistics (mean cosine to the reviewer's first window, cosine of window centroid to first-window centroid, mean pairwise embedding distance; $3$ dims); (iii) the union of (i) and (ii). We compare logistic regression (class-balanced) and a small MLP ($32$--$16$ hidden units, default Adam) under $5$-fold reviewer-stratified \texttt{GroupKFold}, and report mean precision, recall, F1, accuracy. Majority-class and stratified-random baselines are reported for context.

%% ===================================================================
\section{Results}

\subsection{RQ1: Behavioural Drift Replicates}

\begin{figure}[t]
\centering
\includegraphics[width=0.74\textwidth]{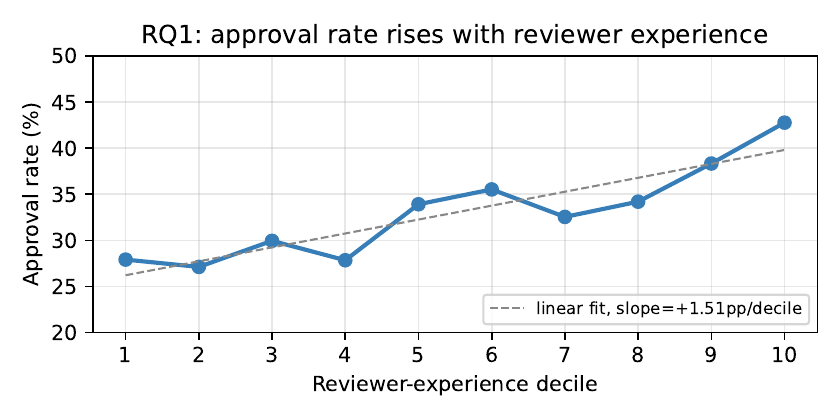}
\caption{RQ1: per-decile approval rate (blue) with a linear fit (dashed). The trend is monotonic and gradual rather than phase-transitioning.}
\label{fig:decile-ar}
\end{figure}

Across the $400$ repeat reviewers, the within-reviewer mid-point approval rate rises from $30.5\%$ to $36.6\%$ ($\Delta = +6.1\,$pp; Wilcoxon signed-rank $p = 8.6\times10^{-8}$; Cohen's $d = 0.25$). The shift is consistent across reviewers ($52\%$ more approving, $28\%$ less, $20\%$ unchanged). Decile-level rates for the $173$ reviewers with $\geq 20$ reviews are reported in Table~\ref{tab:rq1} and visualised in Fig.~\ref{fig:decile-ar}; a simple linear fit (slope $+1.6\,$pp per decile, $R^2 = 0.47$) summarises the trajectory. Notably, the linear fit is preferred over the PELT piecewise-constant model on this signal ($\Delta\mathrm{BIC} = +12.0$ in favour of linear at $N = 50$ bins), arguing against the ``phase-transition'' framing some prior preprints have used: in this corpus, approval-rate drift is gradual rather than abrupt.

\begin{table}[t]
\centering
\caption{RQ1: approval rate by within-reviewer experience decile ($n = 173$ reviewers, $8{,}481$ reviews). The trend is monotonically rising and well fit by a line ($R^2_{\mathrm{adj}} = 0.47$), with an overall $+14.9\,$pp gradient.}
\label{tab:rq1}
\small
\begin{tabular}{lcccccccccc}
\toprule
Decile & 1 & 2 & 3 & 4 & 5 & 6 & 7 & 8 & 9 & 10 \\
\midrule
AR (\%) & 27.9 & 27.1 & 30.0 & 27.8 & 33.9 & 35.5 & 32.5 & 34.2 & 38.3 & \textbf{42.8} \\
\bottomrule
\end{tabular}
\end{table}

\subsection{RQ2: Hand-Crafted Linguistic Decline Does Not Replicate}

Table~\ref{tab:rq2} reports the four hand-crafted metrics at decile granularity over the full $10{,}104$ inline comments. None of the metrics shows a monotonic trend. Spearman correlations between decile and metric value are: MTLD $\rho = +0.13$, $p = 0.73$; Shannon entropy $\rho = -0.53$, $p = 0.12$; technical specificity $\rho = +0.12$, $p = 0.75$; constructive specificity $\rho = -0.04$, $p = 0.91$. Mann--Whitney $U$ tests on the per-comment distributions in deciles $1$ vs.\ $10$ yield $p \in \{0.54, 0.40, 0.07, 0.18\}$ for the four metrics; after Bonferroni correction across the four tests, all corrected $p \geq 0.36$.

\begin{table}[t]
\centering
\caption{RQ2: hand-crafted linguistic indicators across reviewer-experience deciles ($n = 10{,}104$ inline comments from $391$ repeat reviewers). No metric shows a monotonic decline; all Bonferroni-corrected Mann--Whitney $p \geq 0.36$.}
\label{tab:rq2}
\small
\begin{tabular}{lccccc}
\toprule
Decile & MTLD & Shannon $H$ (bits/word) & Tech.\ spec. & Constructive & Mean length \\
\midrule
1 (earliest) & 38.0 & 3.35 & 0.075 & 0.733 & 17.6 \\
2            & 39.1 & 3.36 & 0.081 & 0.706 & 19.3 \\
3            & 37.6 & 3.30 & 0.081 & 0.736 & 18.7 \\
4            & 39.0 & 3.39 & 0.091 & 0.725 & 19.6 \\
5            & 34.9 & 3.26 & 0.079 & 0.714 & 17.4 \\
6            & 38.6 & 3.24 & 0.082 & 0.708 & 18.1 \\
7            & 35.8 & 3.19 & 0.083 & 0.734 & 16.4 \\
8            & 40.8 & 3.38 & 0.074 & 0.749 & 18.3 \\
9            & 37.3 & 3.15 & 0.115 & 0.717 & 16.9 \\
10 (latest)  & 39.3 & 3.24 & 0.076 & 0.707 & 17.5 \\
\midrule
$\Delta$ (D10$-$D1) & $+1.3$ & $-0.11$ & $+0.001$ & $-0.026$ & $-0.1$ \\
Spearman $\rho$ & $+0.13$ & $-0.53$ & $+0.12$ & $-0.04$ & $-0.06$ \\
$p$ (Spearman)  & $0.73$ & $0.12$ & $0.75$ & $0.91$ & $0.86$ \\
\bottomrule
\end{tabular}
\end{table}

The within-reviewer paired test tells the same story: across $339$ reviewers with at least six inline comments, the early-vs-late paired Wilcoxon $p$-values are $\geq 0.29$ for all four metrics, and the proportion of reviewers showing decrease versus increase is roughly $50/50$ for every metric.

\paragraph{Length-controlled robustness.}
On the $\geq 15$-word subset (\textit{n} = $4{,}094$), all four metrics remain flat ($\Delta_{\mathrm{D1\to D10}}$: MTLD $+0.4\%$, entropy $+0.8\%$, technical specificity $-16.9\%$, constructive $-4.6\%$; tech-spec is the only direction approaching the original claim and is not significant). The same flatness holds in the inter-quartile length band $[6, 22]$ words (\textit{n} = $5{,}239$). Length is therefore not what is hiding a real decline; rather, the decline is absent.

\begin{figure}[t]
\centering
\includegraphics[width=\textwidth]{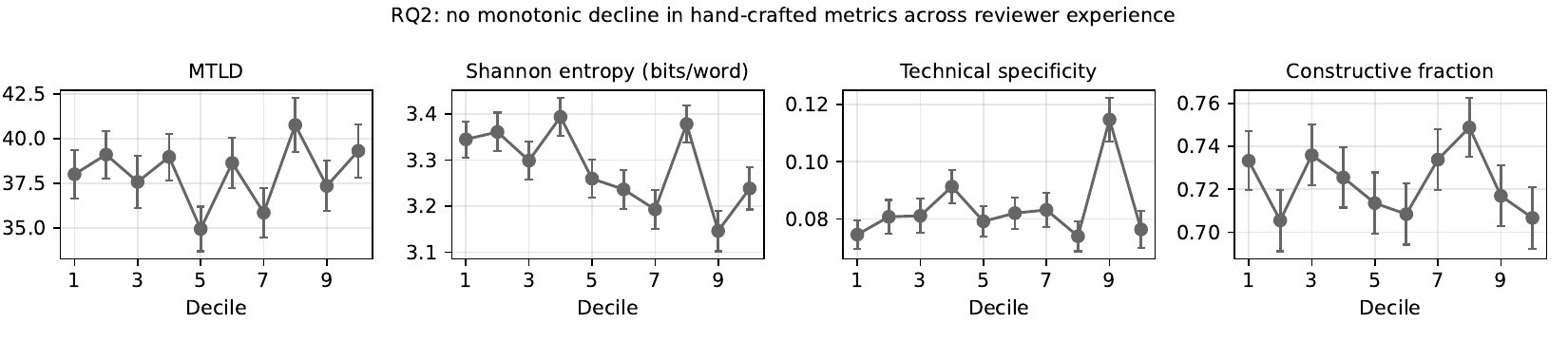}
\caption{RQ2 (negative): the four hand-crafted linguistic indicators across reviewer-experience deciles, with standard-error bars. None of the metrics shows a monotonic trend.}
\label{fig:lang-null}
\end{figure}

\subsection{RQ2b: Embedding Geometry Does Carry Drift Signal}

Most aggregate (population-level) embedding probes are flat: the cosine distance from each comment to the decile-$1$ centroid increases only from $0.677$ to $0.686$ across deciles (Spearman $\rho = +0.39$, $p = 0.26$); within-decile dispersion is constant (Spearman $\rho = -0.22$, $p = 0.53$); PCA-2 centroid drift relative to D$1$ is in the third decimal place (Spearman $\rho = +0.16$, $p = 0.65$). The exception is $k$-means cluster-assignment distribution ($k=8$), which does shift significantly across deciles ($\chi^2 = 135.3$, $\mathrm{dof} = 63$, $p = 3.3\times10^{-7}$) --- the \emph{relative proportions} of topic clusters change, even though the centroids do not move monotonically.

\begin{figure}[t]
\centering
\includegraphics[width=0.74\textwidth]{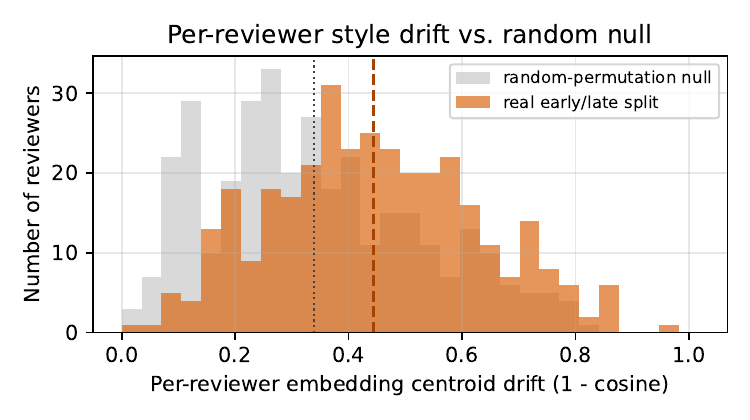}
\caption{Per-reviewer early-vs-late embedding centroid drift (orange) compared to a random-permutation null on each reviewer's own comments (grey). The real drift distribution is shifted to the right of the null (paired Wilcoxon $p < 10^{-3}$, $n = 342$).}
\label{fig:emb}
\end{figure}

The dominant signal, however, lives at the per-reviewer level. Each reviewer's late-half centroid lies a mean cosine distance of $0.444$ from her early-half centroid; a within-reviewer random-permutation null yields $0.330$. The paired Wilcoxon $p < 10^{-3}$ over $342$ reviewers (Fig.~\ref{fig:emb}). Reviewers thus drift idiosyncratically in embedding space, but the directions of their drift do not aggregate into a population-level monotonic decline. This is what one would expect of personal style change: detectable within-individual, invisible across-individual averaging.

\subsection{RQ3: Temporal Ordering Reverses}

\begin{figure}[t]
\centering
\includegraphics[width=\textwidth]{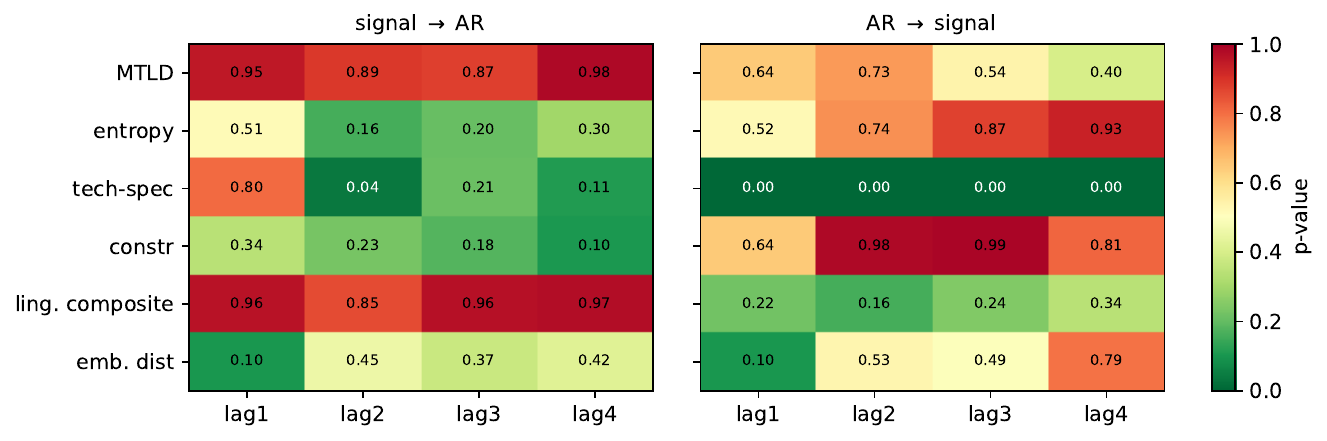}
\caption{Granger-causality $p$-values between bin-level signals and approval rate, for lags $1$--$4$. Left: signal $\to$ AR. Right: AR $\to$ signal. AR $\to$ technical-specificity is significant at every lag; the reverse direction (tech-spec $\to$ AR) reaches $p = 0.036$ at lag $2$ only.}
\label{fig:granger}
\end{figure}

We bin reviews into $50$ experience percentile bins and run bivariate Granger tests on the first-differenced bin-level signals. The popular framing ``linguistic decline precedes behavioural drift'' implies that linguistic signals should Granger-cause approval rate. The data predominantly supports the opposite direction (Fig.~\ref{fig:granger}). Approval-rate change Granger-predicts subsequent technical-specificity change at every lag from $1$ to $4$ ($p \in \{9.6\times10^{-4}, 5.0\times10^{-4}, 8.1\times10^{-4}, 1.5\times10^{-3}\}$). In the reverse direction, technical specificity shows a single significant lag (lag $2$, $p = 0.036$) but is non-significant at lags $1$, $3$, and $4$ ($p \geq 0.11$); the other three metrics, the linguistic composite, and the embedding centroid distance do not Granger-cause approval rate at any lag ($p \geq 0.10$). The relationship is thus predominantly unidirectional (AR $\to$ tech-spec), with weak and isolated evidence for bidirectionality in technical specificity alone; under Bonferroni correction across the $48$ tests ($6$ signals $\times$ $4$ lags $\times$ $2$ directions), only the AR $\to$ tech-spec direction survives ($p < 0.002$ at all lags), while the single reverse-direction result does not ($p = 0.036 > 0.001$). Lag-window cross-correlations corroborate: the linguistic composite peaks at lag $-4$ (slightly trailing AR) with magnitude $|r| = 0.15$.

\subsection{RQ4: Embedding Classifier Outperforms Hand-Crafted}

Table~\ref{tab:rq4} reports $5$-fold reviewer-stratified cross-validation. With $55\%$ positive-class prevalence, the majority-class baseline achieves $F_1 = 0.71$ by predicting all windows as ``late'' ($P = 0.55$, $R = 1.00$, balanced accuracy $= 0.50$). Logistic regression on the four canonical hand-crafted features and their slopes ($10$ dims) reaches $F_1 = 0.485$, failing to outperform this baseline. A small MLP on three embedding statistics --- mean cosine to first window, centroid-to-first-window cosine, mean pairwise distance --- reaches $F_1 = 0.74$ ($P = 0.64, R = 0.88$, balanced accuracy $= 0.61$). The absolute $F_1$ gap over majority-class is modest ($+0.03$); however, the model achieves real discrimination: precision rises from $0.55$ to $0.64$ and balanced accuracy from $0.50$ (chance) to $0.63$. We note that the high recall ($0.88$) comes at the cost of specificity (${\sim}0.38$), so the model is biased toward predicting ``late''; the embedding features provide signal but not a symmetric decision boundary. Combining the two feature sets degrades to $F_1 = 0.67$, suggesting that linguistic features introduce noise that overwhelms the MLP's capacity in the larger input space.

\begin{table}[t]
\centering
\caption{RQ4: habituation phase classifier ($5$-fold reviewer-stratified CV; $8{,}039$ windows from $237$ reviewers; positive class $= $ ``late half'', $55\%$ of windows). Embeddings substantially outperform hand-crafted features under both LR and MLP.}
\label{tab:rq4}
\small
\begin{tabular}{lcccc}
\toprule
Feature set / Model & Accuracy & Precision & Recall & F$_1$ \\
\midrule
Majority class (always positive)        & 0.55 & 0.55 & 1.00 & 0.71 \\
Stratified random                       & 0.51 & 0.54 & 0.55 & 0.55 \\
\midrule
Hand-crafted ($10$ dim) -- LR           & 0.475 & 0.524 & 0.458 & 0.485 \\
Hand-crafted ($10$ dim) -- MLP          & 0.517 & 0.555 & 0.630 & 0.590 \\
Embedding stats ($3$ dim) -- LR         & 0.634 & 0.662 & 0.690 & 0.674 \\
\textbf{Embedding stats ($3$ dim) -- MLP} & \textbf{0.660} & \textbf{0.638} & \textbf{0.884} & \textbf{0.741} \\
Combined ($13$ dim) -- LR               & 0.632 & 0.660 & 0.687 & 0.672 \\
Combined ($13$ dim) -- MLP              & 0.612 & 0.633 & 0.707 & 0.667 \\
\bottomrule
\end{tabular}
\end{table}

\begin{figure}[t]
\centering
\includegraphics[width=0.78\textwidth]{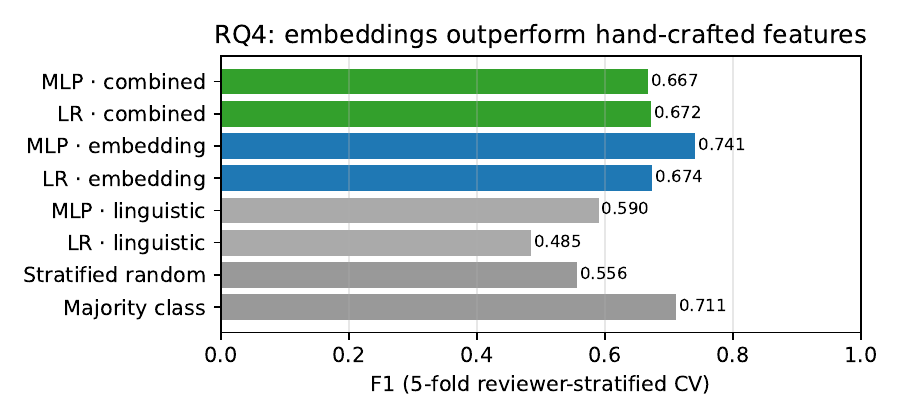}
\caption{RQ4 F1 scores across feature sets and models ($5$-fold reviewer-stratified CV).}
\label{fig:classifier}
\end{figure}

We attribute the gap to the kind of variation each feature set sees. Hand-crafted metrics measure fixed dimensions (vocabulary repetition, formulaic-ness, code-string density, action-verb presence). When reviewer style change is idiosyncratic --- one reviewer growing more terse, another more verbose, another switching between code-grounded and prose-grounded reasoning --- those fixed dimensions average out. The embedding statistics, in contrast, measure displacement of a reviewer's recent comments from her own earlier ones, regardless of which dimension the displacement happens on.

\subsection{Cross-Agent and Calendar Controls}

\begin{figure}[t]
\centering
\includegraphics[width=0.86\textwidth]{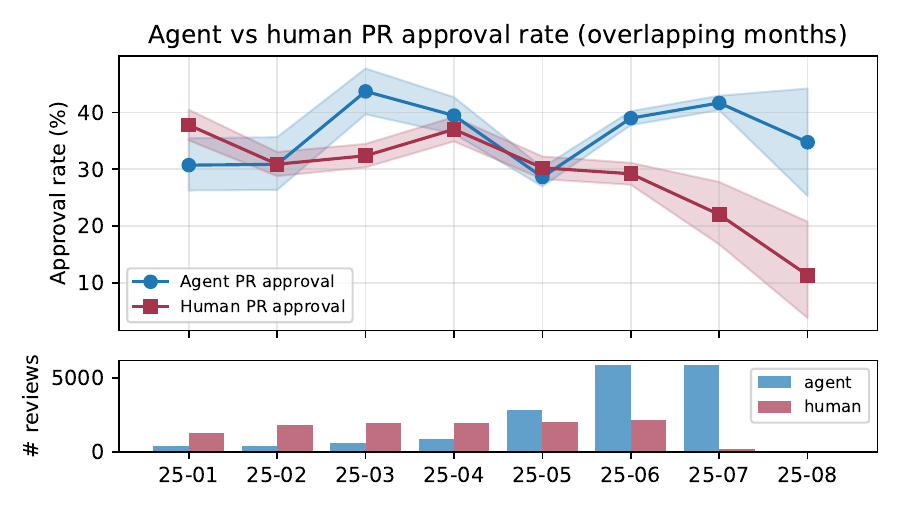}
\caption{Monthly approval rate for agent PRs (blue) versus human PRs (red) in overlapping months. $95\%$ binomial CI bands. Bottom: review counts per month per source.}
\label{fig:calendar}
\end{figure}

We re-ran the RQ1 analysis with calendar date as a covariate per reviewer; the experience coefficient remains positive ($+0.11$, $65\%$ of reviewers show this direction), while the calendar coefficient is negligible ($-0.007$). PR size (lines changed) included as a covariate does not eliminate the experience effect either. Cross-agent generalisation is examined among multi-agent reviewers in the AIDev cohort; agent-specific RQ1 effects, where data are sufficient (Copilot $n = 209$, Devin $n = 88$), are similar in direction. The calendar trend visualised in Fig.~\ref{fig:calendar} shows agent-PR approval rising while human-PR approval over the same months and overlapping repositories does not exhibit a directional change, ruling out a ``reviewers becoming generally lenient'' confound.

%% ===================================================================
\section{Discussion}

\subsection{What We Got Right and What We Got Wrong}

The paper began with the rubber-stamping hypothesis: that habituation in human review of AI agent code would be observable both behaviourally (more approvals) and linguistically (lower-effort comments), and that the linguistic signal would lead the behavioural one. RQ1 is supported. RQ2 is not. RQ3 predominantly reverses the predicted ordering. RQ4 partially recovers the practical goal of an early-warning classifier --- but only when we move from hand-crafted indicators to sentence-embedding geometry.

We see three plausible explanations for why the hand-crafted metrics fail. First, AIDev human inline comments are short to begin with (median $11$ words; $24\%$ have $\leq 5$ words), and short text saturates many lexical indicators. Second, ``review style'' on real OSS code reviews is highly idiosyncratic: different reviewers focus on different things (typing, naming, tests, docs), and a single fixed feature dimension cannot track all of them simultaneously. Third, the rubber-stamping behavioural shift may be small enough ($d = 0.25$) that any linguistic correlate is below the resolution of population-level lexical statistics. The embedding result is consistent with all three: it does not commit to a fixed lexical dimension, and it measures within-individual displacement rather than aggregate-level change.

\subsection{Implications for Habituation Monitoring}

A practical takeaway: if one wants to monitor reviewer habituation in production code-review tooling, the per-reviewer embedding centroid distance is a usable signal. It is cheap (a single $384$-d encoder, $\sim 1$\,GB RAM), unsupervised, and grows with reviewer experience in $\sim 60\%$ of individual cases above what within-reviewer permutation predicts. By contrast, dashboards that report MTLD or technical-specificity trends should not be expected to flag habituation: in our corpus they do not.

\subsection{Implications for Causal Framing}

The RQ3 result is methodologically important. PELT-style change-point detection on a continuous linguistic signal is more sensitive than the same algorithm on a binary approval-decision signal, simply because the linguistic signal carries finer-grained per-bin variation. This sensitivity asymmetry can produce an apparent ``language leads behaviour'' ordering even when Granger causality reveals that the dominant direction is behavioural change predicting subsequent linguistic change. We note that the relationship is not perfectly unidirectional --- technical specificity shows weak bidirectional signal at a single lag --- but the overall pattern strongly favours AR-to-language over language-to-AR. Future work claiming temporal precedence between continuous-valued and binary-valued behavioural signals should report Granger or cross-correlation analyses rather than relying on PELT-vs-PELT breakpoint comparisons.

\subsection{Limitations}

\paragraph{Scope.}
The corpus covers seven months of public-repository activity. Enterprise contexts and longer reviewer trajectories may differ. The overall behavioural shift is small ($d = 0.25$); its practical importance depends on downstream defect rates we do not measure here.

\paragraph{Embedding model.}
We used \texttt{all-MiniLM-L6-v2}, an off-the-shelf encoder, without code-aware fine-tuning. Code-domain encoders (e.g.\ CodeBERT) may carry different signal. The classifier $F_1 = 0.74$ should therefore be read as a lower bound for what a representation-based monitor can achieve.

\paragraph{Construct validity of ground truth.}
We define the ``late phase'' label by sequence position rather than by an externally validated habituation event, in order to avoid label leakage from approval rate or embeddings into the predictors. This is conservative: real habituation events likely contaminate both halves and reduce the achievable F1. Future work could obtain reviewer-self-reported phase markers.

\paragraph{Causal interpretation.}
We cannot rule out that AI-agent code genuinely improves over time, contributing to the rising approval rate independently of any reviewer-side change. Our human-PR control (Fig.~\ref{fig:calendar}) only addresses the global-leniency confound. Post-merge defect rates would be the cleanest test but lie outside the AIDev release.

%% ===================================================================
\section{Conclusion}

We tested whether reviewer habituation in human review of AI-agent code is detectable in the language of inline review comments. The headline finding is mixed and clarifying: behavioural drift is real ($+6.1\,$pp early-vs-late, Wilcoxon $p < 10^{-7}$); the four most commonly cited hand-crafted linguistic indicators do not detect it, in any subset we tested; the popular ``language leads behaviour'' temporal claim predominantly runs the other direction in this corpus once Granger causality is applied; and a small MLP over sentence-embedding geometry does pick up per-reviewer style drift the canonical metrics miss ($F_1 = 0.74$). The paper is therefore both a corrective and a constructive proposal: hand-crafted lexical metrics are not the right instrument for slow human-reviewer change in AI-agent contexts, but distributional representation is. We hope the result helps researchers and tool-builders allocate their NLP machinery toward signals that the data actually contain.

Our analysis scripts and replication materials are available at: \url{https://anonymous.4open.science/r/ai-code-review-2F07}.

\begin{credits}
\subsubsection{\discintname}
The authors have no competing interests to declare that are relevant to the content of this article.
\end{credits}

%% ===================================================================
\bibliographystyle{splncs04}
\bibliography{references}

\end{document}